\documentclass[
    ,final            
  ]
  {aipproc}

\layoutstyle{8x11double}

\usepackage{amsmath}

\newcommand{\Dsl}{{\slash \negthinspace \negthinspace \negthinspace \negthinspace  D}}

\begin{document}

\title{Fate of Yang-Mills black hole in early Universe}


\author{\L ukasz Nakonieczny \footnote{lnakonieczny@kft.umcs.lublin.pl}}{
  address={Institute of Physics \protect \\
Maria Curie-Sk{\l}odowska University \protect \\
20-031 Lublin, pl.~Marii Curie-Sk{\l}odowskiej 1, Poland }
}

\author{Marek Rogatko \footnote{rogat@kft.umcs.lublin.pl}}{
  address={Institute of Physics \protect \\
Maria Curie-Sk{\l}odowska University \protect \\
20-031 Lublin, pl.~Marii Curie-Sk{\l}odowskiej 1, Poland }
}

\begin{abstract}
According to the Big Bang Theory as we go back in time the Universe becomes progressively hotter
and~denser. This leads us to believe that the early Universe was filled with hot plasma of elementary
particles. Among many questions concerning this phase of history of the Universe there are questions
of existence and fate of magnetic monopoles and~primordial black holes.
Static solution of Einstein-Yang-Mills system may be used as a toy model for such a black
hole. Using methods of field theory we will show that its existence and regularity depend
crucially on the presence of fermions around it. 
\end{abstract}

\maketitle


\section{Introduction}

Recent observational data strongly suggest the existence of dark matter (DM) \cite{WMAP-7}.
One of a few remaining DM candidates compatible with the Standard Model (SM) are primordial black holes (PBH) \cite{Frampton}.
These black holes were formed in the early Universe during gravitational collapse of density fluctuation. 
The existence of the PBH was first proposed in \cite{Zeldowich, Havking,Carr1} and their formation during inflation or
phase transition in the early Universe was discussed in \cite{Kholpov1,Carr2,Kholpov2}. 
Another type of objects that could be formed during phase transition in early Universe are topological defects like
cosmic strings and magnetic monopoles \cite{vilenkin}. If these objects really formed during evolution of the Universe
we should consider the possibility of interactions among them. Within the framework of field theory a simple model of the system that 
allows the existence of magnetic monopoles may be given by Yang-Mills (YM) theory with $SU(2)$ gauge group. 
Particle like solutions of Einstein-Yang-Mills (EYM) system describing the magnetic monopole were first constructed in \cite{Bartnik}.
On the other hand, solutions describing a black hole with magnetic monopole in this context were presented in \cite{Bizon1}.
There was also shown in \cite{Bizon2} that finiteness of the black hole mass demands absence of an electric part of YM field.

Another important component of the early Universe whose interaction with PBH we should consider is fermionic matter.
To describe these interactions we use methods of field theory. As a toy model of PBH we take the aforementioned 
spherically symmetric YM black hole. As a representation of fermionic matter we use Dirac field. 
Using dimensional reduction and~bosonization we will analyze the influence of fermions on the considered PBH.

\section{Dirac equation in magnetic Yang-Mills black hole background}

A general form of a spherically symmetric and time independent metric is
\begin{align}
ds^2 = - A^2(r) dt^2 + B^{-2}(r)dr^2 + r^2 d\Omega^2,
\end{align}
where $r$ is an usual radial coordinate and $d\Omega^2 = d\theta^2 + \sin^2(\theta) d\phi^2$
is a standard metric on two dimensional sphere. 
The location of an event horizon of a black hole is given by the largest positive root of $A^2(r)$, 
which is denoted as $r_{H}$. Actually for analyzing fermions equations of motion it is more convenient 
to introduce the so-called {\it tortoise } coordinate $r_{*}$. This coordinate is given by the following
relation:
\begin{align}
dr_{*} = \frac{B(r)}{A(r)} dr. 
\end{align}
Our metric expressed in coordinates ($t,r_{*},\theta,\phi$) is
\begin{align}
ds^2 = A^2( - dt^2 + dr_{*}^2) + C^2 d\Omega^2,
\end{align}
where we introduce a function $C$ to keep in mind that now we have $r = r(r_{*})$.
From now on, to simplify our notation we will omit explicit writing of arguments of functions.
During our computations we will use tetrad formalism. The local ortogonal tetrad is defined by the relation
$g_{\mu \nu} = e^{i}_{\mu} e^{j}_{\mu} \eta_{i j}$, where $e^{i}_{\mu}$ is an element of the tetrad, 
$g_{\mu \nu}$ and $\eta_{i j}$ are metric tensors in curved and flat spacetimes respectively.
We use the following signature convention: $\eta_{00} = -1$, $\eta_{11} = \eta_{22} = \eta_{33} = + 1$. 

The general form of static and spherically symmetric Yang-Mills potential may be written as
\begin{align}
H_{\mu} = e^{i}_{\mu} [ a_{i} n^{k} \tau_{k} + \frac{ 1 - w(r)}{ 2 \lambda C} \varepsilon_{ i j k} n^{j} \tau^k], 
\end{align}
where $a_{i} = (a_{0}, a_{1})$ and $w$ represent electric and magnetic parts of YM field,
$n^{i}$ is a unit vector normal to the sphere, $\lambda$ is a coupling constant, $\varepsilon_{i j k}$ stands for 
a totally anti-symmetric Levi-Civita symbol, and $\tau^k$ are generators of $SU(2)$ gauge group represented
by Pauli matrices. 
As the representation for four dimensional gamma matrices we choose Weyl basis
\begin{align}
\gamma^0 = \begin{bmatrix} 0& I \\ I & 0 \end{bmatrix}, \qquad
\gamma^{i} = \begin{bmatrix} 0&\sigma^{i}  \\ -\sigma^i & 0 \end{bmatrix},
\end{align}
where $I$ is a unit two dimensional matrix and $\sigma^{i}$ are three Pauli matrices.
Gamma matrices in curved spacetime are given by the standard relation: $\gamma^{\mu} = e^{\mu}_{k} \gamma^k$.
Representing a~spinor $\psi$ as $\psi = \begin{pmatrix} \psi_{L} \\ \psi_{R} \end{pmatrix}$ we may 
write the Dirac equation as
\begin{align}
\label{system_4d}
i \Dsl^{+} \psi_{R} - m \psi_{L} = 0, \nonumber \\
i \Dsl^{-} \psi_{L} - m \psi_{R} = 0,
\end{align}
where operators $\Dsl^{\pm}$ are given by \cite{Gibbons,LN1}
\begin{align}
\Dsl^{\pm} &= A^{-1}\partial_{t} - 
i \lambda \{ [ \sigma^{0}a_{0} \pm \sigma^{1}a_{1} ] \bar{n} \cdot \bar{\tau} \pm 
\frac{ w - 1}{ 2 \lambda C } \bar{n} \cdot \bar{\sigma} \times \bar{\tau} \} + \nonumber \\
&\pm  \bar{\sigma} \cdot \bar{n} A^{-1}\partial_{r_{*}} 
\pm  \bar{\sigma} \cdot \bar{n} \{ A^{-1} C^{-1} \partial_{r_{*}} C +  
\frac{ 1 }{ 2 } A^{-1} A^{-1} \partial_{r_{*} }A  \} + \nonumber \\
&\pm  C^{-1} D_{S^2}. 
\end{align}
In the above formula a bar over a quantity represents three dimensional vector, 
a dot -- scalar multiplication,
$\times$ -- vector multiplication, and $D_{S^2}$ is Dirac operator on $S^2$.
We are mainly interested in fermions in {\it s-wave} sector, which is described by the lowest eigenvalue of
$D_{S^2}$. The action of operators $\bar{n} \cdot \bar{\tau}$, $\bar{n} \cdot \bar{\sigma} \times \bar{\tau}$,
$\bar{\sigma} \cdot \bar{n}$ and $D_{S^2}$ on these states may be found in \cite{Gibbons}.
We may use this knowledge to integrate over the angles in four dimensional Dirac action to obtain effective 
two dimensional theory in ($t,r_{*}$) plane. 

On the other hand, dimensional reduction of the YM action gives us the following Lagrangian:
\begin{align}
L_{YM-2d} = - \frac{C^2}{4} f_{a b}f^{a b} - |d w^2| - \frac{1}{2C^2}(|w|^2 - 1)^2,
\end{align} 
where $d_{a} = \nabla_{a} - i B_{a}$, $f_{a b} = \partial_{a} B_{b} - \partial_{b} B_{a}$, and $B_{a} = e^{i}_{a} a_{i}$.
From now on latin letters from the beginning of alphabet will label indexes from curved two dimensional ($t,r_{*}$) 
spacetime and those from the middle from flat spacetime.

\section{Massless fermions}

In the massles case we see from (\ref{system_4d}) that chiralities decouple and effective two dimensional theory
contains two fermionic fields connected to $\psi_{L}$ and $\psi_{R}$. 
The effective Lagrangian for the field $\psi_{R}$ is
\begin{align}
\label{FR-2d}
\mathcal{L}_{FR-2d} = - i \bar{G}_{R} \tilde{\gamma}^{a} \nabla_{a} G_{R} - \lambda  B_{a} 
\bar{G}_{R} \tilde{\gamma}^{a} \tilde{\gamma}^3 G_{R} + \nonumber \\ 
+ V \bar{G}_{R} \tilde{\gamma}_{L} G_{R} - V \bar{G}_{R} \tilde{\gamma}_{R} G_{R},
\end{align}  
where $V = \frac{w}{C}$.
In this formula $G_{R}$ is two dimensional spinor field connected to $\psi_{R}$ by rescaling 
and multiplication by appropriate $\sigma$ matrices \cite{LN2}.
Two dimensional flat spacetime gamma matrices are given by the following relations:
\begin{align}
&\{ \tilde{\gamma}^i , \tilde{\gamma}^j \} = 2 \eta^{i j} ,\qquad \eta_{00} = -1 = - \eta_{11}, \nonumber \\
&\tilde{\gamma}^0 = - i \sigma^3, \qquad \tilde{\gamma}^1 = - \sigma^2, \nonumber \\ 
&\tilde{\gamma}^3 = \tilde{\gamma}^0 \tilde{\gamma}^1 = \sigma^1, \qquad
\tilde{\gamma}_{L/R} = \frac { 1 }{ 2 }( I \pm \tilde{\gamma}^3 ).
\end{align}
The Lagrangian for the field connected to $\psi_{L}$ differs from (\ref{FR-2d}) only by a sign of a term proportional to $B_{a}$. 

Analyzing equations of motion for fermions in curved spacetime is a highly nontrivial and difficult task. However, in case at 
hand since we can express our problem in form of effective two dimensional theory we may use bosonization 
technique \cite{Zinn-Justin}. 
This technique is well defined for flat spacetime two dimensional problems and~also for asymptotically flat spacetimes
\cite{Gamboa,Barcelos-Neto}. 
The~merit of bosonization is that we express a fermionic sector of our theory in terms of complex scalar field. 
Basic bosonization formulas in our case are given below
\begin{align}
\label{bosonization}
&j^{a} = \bar{\psi} \tilde{\gamma}^a \psi  = \frac{1}{ \sqrt{\pi}} \varepsilon^{a b}\nabla_{b} \phi,  \\ 
&j^{3 a} = \bar{\psi} \tilde{\gamma}^a \gamma^3 \psi = \frac{1} {\sqrt{\pi}} \nabla^{a} \phi, \\
&\bar{\psi} \gamma_{L} \psi = b e^{ 2 i \sqrt{\pi} \phi}, \qquad 
&\bar{\psi} \gamma_{R} \psi = b e^{ - 2 i \sqrt{\pi} \phi}.
\end{align} 

After bosonization from (\ref{FR-2d}) we obtain the following scalar Lagrangian:
\begin{align}
\mathcal{L}_{BR} = - \frac{1}{2} \nabla_{a}\phi_{R} \nabla^{a} \phi_{R} - \lambda B_{a} 
\frac{1}{\sqrt{\pi}} \nabla^a \phi_{R} + \nonumber \\
+ V b ( e^{ 2 i \sqrt{\pi} \phi_{R}}  - e^{ - 2 i \sqrt{\pi} \phi_{R}} ), 
\end{align}
from which we derive an equation of motion for $\phi_{R}$ field
\begin{align}
\label{sclar_massless_R}
\nabla_{a} \nabla^{a} \phi_{R} +  \frac{ \lambda }{ \sqrt{\pi} } \nabla_{a} B^{a} 
+ 4 i b \sqrt{\pi} V \cos( 2 \sqrt{\pi} \phi_{R}) = 0.
\end{align}
Then, we derive an equation of motion for scalar field connected to the $\psi_{L}$ sector 
of our original fermionic theory analogically:
\begin{align}
\label{sclar_massless_L}
\nabla_{a} \nabla^{a} \phi_{L} -  \frac{ \lambda }{ \sqrt{\pi} } \nabla_{a} B^{a} 
+ 4 i b \sqrt{\pi} V \cos( 2 \sqrt{\pi} \phi_{L}) = 0.
\end{align}

Equations (\ref{sclar_massless_R}) and (\ref{sclar_massless_L}) are highly nonlinear and we were not able to
find a solution in terms of known special functions in the whole spacetime. Nevertheless, we may find some useful 
information about their solutions by analyzing them in two asymptotic regions, namely in near horizon region 
and in large $r_{*}$ region. 

In the large $r_{*}$ region we have that
\begin{align}
A^2 \approx 1, \qquad C = r, \qquad w \approx \pm 1 ,\qquad r_{*} \sim r.
\end{align} 
Taking this into account the equation (\ref{sclar_massless_R}) turns into
\begin{align}
- \partial_{t}^2 \phi_{R} + \partial^2_{r_{*}} \phi_{R}  +
4 i b \sqrt{\pi} \frac{w(\infty)} {r} \cos( 2 \sqrt{\pi} \phi_{R}) = 0.
\end{align}
After dropping a term proportional to $O(r^{-1})$ we get a~free wave equation
\begin{align}
- \partial_{t}^2 \phi_{R} + \partial^2_{r_{*}} \phi_{R}  =0.
\end{align}
A regular solution to this equation may be expressed as a plane wave
\begin{align}
\phi_{R} = c_0  e^{ - i \omega (t \pm r_{*})},
\end{align}
where $c_0$ is an integration constant. 
On the other hand, in~the near horizon region we have
\begin{align}
A^2 = 2 \kappa (r - r_{H}), \quad C = r_{H}, \quad r - r_{h} = e^{2 \kappa r_{*}},
\end{align}
where $\kappa$ is surface gravity of a Yang-Mills black hole.
The equation (\ref{sclar_massless_R}) in this region takes the form
\begin{align}
- \partial_{t}^2 \phi_{R} +  \partial_{r_{*}}^2 \phi_{R}  + 
 i \frac{ 8 b \sqrt{\pi} w(r_{h}) \kappa }{ r_{h} } e^{2 \kappa r_{*}}
 \cos( 2 \sqrt{\pi} \phi_{R} ) = 0.
\end{align}
Because as we approach a black hole horizon $r_{*} \rightarrow - \infty$, we
may drop a term proportional to $e^{2 \kappa r_{*}}$ in the above equation. From this we see that 
a regular solution is again a time dependent plane wave given by
\begin{align}
\phi_{R} = c_1 e^{ - i \omega ( t \pm r_{*} ) },
\end{align}
where $c_{1}$ is some other integration constant. 
On the basis of this analysis we may conclude that a regular solution to 
the considered scalar field equation will be time dependent. But both scalar fields 
represent fermionic currents and their time dependence ultimately means that 
fermionic fields will also be time dependent. 
To see how this may influence YM field we use equations of motion for electric and magnetic
parts of this field in the presence of fermions. After using bosonization formulas
(\ref{bosonization}) these equations read \cite{LN2}
\begin{align}
\label{massless_B_phi}
&\nabla_{a} [ C^2 f^{a b} ] - 2~ |w|^2 B^{b} = \frac{\lambda}{\sqrt{\pi}}[\nabla^b \phi_{R} - \nabla^b \phi_{L}],\\
&\nabla_{a} \nabla^{a} w - \frac{ 2 }{ C^2 } w ( |w|^2 - 1 ) + 2 w B_{a}B^{a} = \nonumber \\
&\qquad i \frac{ 2 b }{ C }~ [ \sin( 2 \sqrt{\pi} \phi_{R}) + \sin(2 \sqrt{\pi} \phi_{L} ) ].
\label{massless_w_phi}
\end{align}
 
Conclusions that stem from the above equations are the following.
First, let us remind that equations (\ref{sclar_massless_R}) and~(\ref{sclar_massless_L}) differ only by sign in front of the term 
proportional to $B_{a}$ and we assume that initially a black hole has only magnetic charge ($B_{a} = 0$).  
Second, from equation (\ref{massless_B_phi}) we see that in this case contributions of scalar fields cancel each other 
and $B_{a} = 0$ is still a valid solution to (\ref{massless_B_phi}).  
Third, from equation (\ref{massless_w_phi}) we see that there is a nonzero contribution of scalar fields to magnetic
part of YM field. But since these fields are time dependent so should be the resulting $w$. On the other 
hand, through Einstein equations, this results in time dependence of metric tensor elements and ultimately means
that the assumption of staticity of Yang-Mills black hole is destroyed in the presence of massless fermions.

\section{Massive fermions}

For massive fermions we see from (\ref{system_4d}) that chiralities are mixed up. To use bosonization in
this case we need to make an additional assumption about the form of $\psi_{R}$ and~$\psi_{L}$. We use the 
following ansatz:
\begin{align}
G_{L} = i \sigma^3 G_{R} \equiv G, 
\end{align}
where $G_{L}$ and $G_{R}$ are two dimensional spinor fields connected with $\psi_{L}$ and $\psi_{R}$ respectively \cite{LN2}.
Having this in mind an effective two dimensional fermionic Lagrangian is as follows:
\begin{align}
\label{fermion-2d-massive}
\mathcal{L}_{GF-2d} = - i \bar{G} \tilde{\gamma}^a \nabla_{a} G 
-  \lambda B_{a}  \bar{G} \tilde{\gamma}^a \tilde{\gamma}^3 G + \nonumber \\
+ ( V + m ) \bar{G} \tilde{\gamma}_{L} G + ( m - V) \bar{G} \tilde{\gamma}_{R} G, 
\end{align}
where like in massless case $V = \frac{w}{C}$.
Using the same bosonization formulas as in the massless case we arrive at the following scalar Lagrangian:
\begin{align}
\mathcal{L}_{G B} &= - \frac{1}{2} \nabla_{a} \phi \nabla^a \phi - \frac{ \lambda }{ \sqrt{\pi} } B_{a}
 \nabla^a \phi + \nonumber \\
&+ ( V + m ) b e^{ 2 i \sqrt{\pi} \phi} + ( m - V ) b e^{ - 2 i \sqrt{\pi} \phi}.
\end{align}
The equation of motion for scalar field in this case is
\begin{align}
\label{sclar_massive}
&\nabla_{a} \nabla^a \phi + \frac{ \lambda }{ \sqrt{\pi} } \nabla_{a} B^a + \nonumber \\
&+ 4 i b\sqrt{\pi} 
\bigg \{ 
V \cos( 2 \sqrt{\pi} \phi ) + i m \sin( 2 \sqrt{\pi} \phi)
\bigg \} = 0. 
\end{align}
In the large $r_{*}$ limit, after dropping terms proportional to $O(r^{-1})$, we obtain a sine-Gordon equation
\begin{align}
- \partial_{t}^2 \phi + \partial^2_{r_{*}} \phi - 
4 b \sqrt{\pi} m \sin( 2 \sqrt{\pi} \phi) = 0,
\end{align}
for which a regular time dependent and decaying at infinity (kink type) solution is given by
\begin{align}
\phi = \frac{ 2 }{ \sqrt{\pi} } \arctan \bigg( e^{ - \sqrt{ \frac{ 8 b \pi m }{ 1 - v^2 }  }( r_{*} - vt )  } 
\bigg).
\end{align}
On the other hand, the same type of analysis like in the massless case revealed that in the near horizon limit we 
again have a free wave equation with a regular time dependent solution in form of a plane wave:
\begin{align}
\phi = c_2 e^{ - i \omega ( t \pm r_{*} ) }. 
\end{align}

Now we will discuss the influence of massive fermions on a YM black hole.
Yang-Mills equations of motion in the presence of fermions (after bosonization) are the following:
\begin{align}
\label{ym_massive-1}
&\nabla_{a} [ C^2 f^{a b} ] - 2~ |w|^2 B^{b} = \frac{\lambda}{\sqrt{\pi}} \nabla^b \phi,\\
&\nabla_{a} \nabla^{a} w - \frac{ 2 }{ C^2 } w ( |w|^2 - 1 ) + 2 w B_{a}B^{a} = \nonumber \\
&\qquad i \frac{ 2 b }{ C } \sin( 2 \sqrt{\pi} \phi). 
\label{ym_massive-2}
\end{align}
From equation (\ref{ym_massive-1}) we see that, even if we initially set $B_{a} = 0$, fermions
give a nonzero contribution to the electric part of YM field. This means that contrary to the massless case the
presence of fermions leads to dyonic structure of a black hole.
But as was shown in \cite{Bizon2} dyonic Yang-Mills black hole necessarily has infinite mass.
On the other hand, form equation (\ref{ym_massive-2}) we see that time dependent fermions give a nonzero contribution to the
magnetic part of YM field. Through Einstein equations this leads to a time dependent line element and the
destruction of staticity of a black hole.

\section{Conclusions}

Now we shall present a short summary of our results concerning the influence of fermions on a 
primordial black hole modeled by a magnetic Yang-Mills black hole.

First we will discuss the massless case. By asymptotic analysis we find an evidence that regular solutions
to equations (\ref{sclar_massless_R}) and (\ref{sclar_massless_L}) should be time dependent.
These equations describe bosonized massless fermions and give a nonzero contribution to a magnetic part of
Yang-Mills field. Because their solutions are time dependent the resulting magnetic part of YM field will be
time dependent and, through Einstein equations, elements of a metric tensor will also be time dependent. 

In massive case bosonized fermions are described by solutions to equation (\ref{sclar_massive}).
The same type of analysis like in massless case also reveals the destruction of staticity of our black hole.
Moreover, massive fermions will lead to the formation of a dyonic black hole. But, as was shown in \cite{Bizon2}, the presence
of an electric part of Yang-Mills field leads to infinite mass of the resulting black hole. 

In conclusion we may say that the presence of fermions and their interaction with the considered PBH
will lead to its destruction through mechanisms described above.

\end{document}